\documentclass[aps, prd, twocolumn, nofootinbib, 
floatfix,superscriptaddress]{revtex4}
\usepackage{epsfig}
\usepackage{amsmath}
\usepackage{amsfonts}
\usepackage{amssymb}
\usepackage{amssymb,amsmath,amsfonts}
\usepackage{xfrac}
\usepackage{color}
\usepackage[utf8]{inputenc}
\usepackage{graphicx}
\usepackage{dcolumn}
\usepackage{bm}
\usepackage{tikz}
\usepackage{amssymb}
\usepackage{float}
\usepackage{amsmath}
\usepackage{dcolumn}
\usepackage{cancel}

\usepackage{tcolorbox}
\usepackage{hyperref}
\usepackage{booktabs}
\hypersetup{colorlinks, citecolor=red, linkcolor=bluscuro, urlcolor=bluscuro}
\definecolor{rossos}{cmyk}{0,1,1,0.55}
\definecolor{bluscuro}{rgb}{0.15, 0.2, .85}
\definecolor{bluchiaro}{cmyk}{1,.3,0.,0.1}

\newcommand{\mpl}{M_{\mathrm{Pl}}}

\newcommand{\Omf}{\Omega_\mathrm{f}}
\newcommand{\Omr}{\Omega_\mathrm{r}}
\newcommand{\rh}{\rho}
\newcommand{\pf}{p_\mathrm{f}}
\newcommand{\rhof}{\rho_\mathrm{f}}
 
\newcommand{\wf}{w_\mathrm{f}}
\newcommand{\wtot}{w_{\mathrm{tot}}}
\newcommand{\Hh}{H}
\newcommand{\dd}{\mathrm{d}}

\def\doi{http://doi.org}

\begin{document}

\title{Primordial black hole formation from transient 
$f(T)$ cosmology }

\author{Gerasimos Kouniatalis}
\email{gkouniatalis@noa.gr}
 \affiliation{National Observatory of Athens, Lofos Nymfon, 11852 Athens, 
Greece}
\affiliation{Physics Department, National Technical University of Athens,
15780 Zografou Campus,  Athens, Greece}

\author{Theodoros Papanikolaou}
\email{papaniko@upatras.gr}
\affiliation{Department of Physics, University of Patras, 26504, Patras, Greece}
\affiliation{Scuola Superiore Meridionale, Largo San Marcellino 10, 80138 Napoli, Italy}
\affiliation{Istituto Nazionale di Fisica Nucleare (INFN), Sezione di Napoli, Via Cinthia 21, 80126 Napoli, Italy}
 \affiliation{National Observatory of Athens, Lofos Nymfon, 11852 Athens, 
Greece}

\author{Spyros Basilakos}
\email{svasil@academyofathens.gr}
\affiliation{Academy of Athens, Research Center for Astronomy $\&$ Applied 
Mathematics, Soranou Efessiou 4, 11-527, Athens, Greece }
\affiliation{National Observatory of Athens, Lofos Nymfon, 11852 Athens, Greece}
\affiliation{School of Sciences, European University Cyprus, Diogenes Street, Engomi, 1516 Nicosia, Cyprus}

\author{Emmanuel N. Saridakis} \email{msaridak@noa.gr}
\affiliation{National Observatory of Athens, Lofos Nymfon, 11852 Athens, Greece}
\affiliation{CAS Key Laboratory for Researches in Galaxies and Cosmology, 
School 
of Astronomy and Space Science, University of Science and Technology of China, 
Hefei, Anhui 230026, China}
\affiliation{Departamento de Matem\'{a}ticas, Universidad Cat\'{o}lica del 
Norte, Avda. Angamos 0610, Casilla 1280 Antofagasta, Chile}


\begin{abstract}

We study primordial black hole (PBH) formation in a minimally coupled $f(T)$ 
teleparallel cosmology that generates a transient departure from standard 
radiation domination. The model is constructed so that modified-gravity effects 
are negligible at early and late times, but become dynamically relevant over a 
finite epoch, during which an effective torsion component reduces the total 
equation-of-state parameter below $1/3$.
We show that this transient softening lowers the collapse threshold for 
overdensities at horizon re-entry, leading to an exponential enhancement of PBH 
formation. In addition, the modified background alters the relation between 
temperature and horizon mass, producing a localized feature in the PBH mass 
function. 
For representative parameters, PBHs with asteroid-scale masses can account for 
a significant fraction, or even the entirety, of dark matter for perturbation 
amplitudes $\sigma^2 \sim \mathcal{O}(10^{-3})$, while remaining consistent with 
current constraints. Our results demonstrate that modified gravity alone can 
efficiently generate PBHs, without requiring ad hoc modifications of the 
radiation sector.

\end{abstract}

\maketitle

\section{Introduction}

The possibility that black holes may have formed in the early Universe, before 
the onset of structure formation, provides a conceptually economical probe of 
physics at energy and curvature scales far beyond those accessible through 
late-time astrophysical 
observations~\cite{Chapline:1975ojl,Khlopov:1985fch,Ivanov:1994pa}. Primordial 
black holes (PBHs) can form when sufficiently large curvature or density 
perturbations re-enter the Hubble horizon and undergo gravitational collapse, 
converting a fraction of the cosmic energy density into compact objects whose 
masses are determined, up to an efficiency factor, by the horizon mass at the 
time of formation~\cite{1971MNRAS.152...75H,Carr:1974nx,Carr:1975qj}. Since the 
PBH formation probability is extremely sensitive both to the statistical 
properties of primordial perturbations and to the microphysics of the cosmic 
medium, PBHs offer a powerful framework in which one can connect inflationary 
physics~\cite{Starobinsky:1980te,Guth:1980zm,Linde:1981mu,Albrecht:1982wi,
Linde:1983gd} with gravitational dynamics beyond general 
relativity \cite{Clifton:2011jh,Capozziello:2011et}, and have been studied in 
detail 
\cite{Clesse:2015wea,Clesse:2016vqa,Ali-Haimoud:2016mbv,
Niikura:2017zjd,Kalaja:2019uju,Niikura:2019kqi,
Carr:2020gox,
Ai:2024cka,
Montefalcone:2025akm,Wang:2025dbj,Carr:2026hot}. 

In the standard radiation-dominated (RD) scenario, characterized by the 
relativistic equation of state (EoS) $p=\rho/3$, pressure gradients act 
efficiently against the collapse of overdense regions. Numerical simulations and 
analytic arguments indicate that only perturbations exceeding a relatively high 
threshold amplitude at horizon re-entry can form 
PBHs~\cite{Musco:2004ak,Musco:2012au}. Hence, even modest changes in the 
collapse threshold can induce exponentially large variations in the initial PBH 
mass fraction when the primordial fluctuations are approximately 
Gaussian~\cite{Sasaki:2018dmp}. It is therefore natural to investigate 
mechanisms that can temporarily reduce the efficiency of pressure gradients. A 
characteristic example is provided by the quantum chromodynamics (QCD) epoch, 
during which the effective number of relativistic degrees of freedom changes and 
the EoS, as well as the sound speed, depart from the ideal relativistic value. 
In that case the cosmic fluid becomes transiently softer, the collapse threshold 
decreases, and PBH formation is enhanced on the mass scales associated with the 
horizon mass at that epoch~\cite{Jedamzik:1996mr,Musco:2023dak}, namely around 
the solar-mass range. More generally, any physical mechanism that induces a 
time-localized reduction of the effective EoS or sound speed can imprint a 
localized feature in the PBH mass function.

On the other hand,   modified theories of gravity have been 
extensively investigated, motivated both by 
theoretical considerations and by cosmological 
observations
\cite{CANTATA:2021asi,  Nojiri:2017ncd}. 
In particular, one may start from the standard curvature-based 
formulation and extend the Einstein-Hilbert Lagrangian in various ways, leading 
for instance to $f(R)$ 
gravity \cite{Capozziello:2002rd,DeFelice:2010aj,
Nojiri:2010wj}, $f(G)$ gravity~\cite{Nojiri:2005jg,DeFelice:2008wz}, 
higher-order or cubic curvature corrections \cite{Asimakis:2022mbe}, and 
Lovelock gravity \cite{Lovelock:1971yv,Deruelle:1989fj}. 
Alternatively, one may consider equivalent formulations of gravity based on 
different geometric quantities. Specifically, starting from the torsional 
formulation of gravity, namely the Teleparallel Equivalent of General 
Relativity, one is led to extensions such as $f(T)$ 
gravity \cite{Cai:2015emx,Linder:2010py}, $f(T,T_G)$ 
gravity \cite{Kofinas:2014owa,Kofinas:2014daa}, and $f(T,B)$ 
gravity \cite{Bahamonde:2015zma,Bahamonde:2016grb}. Such torsional 
modifications of gravity have been shown to lead to interesting cosmological 
phenomenology  
\cite{Chen:2010va,Bengochea:2010sg,Cardone:2012xq, Cai:2011tc, 
Otalora:2014aoa,
Cai:2018rzd,Cai:2019bdh, 
Golovnev:2020las, Caruana:2020szx,
Mavromatos:2021hai,  
Capozziello:2022dle,Ren:2022aeo, Yang:2024kdo,Bouhmadi-Lopez:2026dte}.

The purpose of the present work is to investigate a class of $f(T)$ 
gravity models in which the deviation from the teleparallel equivalent of 
general relativity is explicitly transient. In particular, we construct a viable 
$f(T)$ scenario in which the modification becomes negligible both in the 
asymptotic early-time regime, where the torsion scalar is large, and in the 
late-time regime, where the torsion scalar tends to zero, while it becomes 
dynamically relevant over a finite intermediate epoch. During this period, the 
torsion contribution can be interpreted as an effective fluid component with a 
time-dependent EoS parameter. By an appropriate choice of model parameters, the 
effective torsion sector constitutes a non-negligible fraction of the total 
energy density for a finite interval, while its pressure becomes negative. As a 
consequence, the total EoS parameter 
$w_{\mathrm{tot}}=p_{\mathrm{tot}}/\rho_{\mathrm{tot}}$ is temporarily reduced 
below the canonical radiation value $1/3$.

This transient softening has direct implications for PBH formation. Since the 
collapse threshold at horizon re-entry depends on the background EoS, and more 
precisely on the efficiency with which pressure gradients can counterbalance 
gravity, a temporary reduction of $w_{\mathrm{tot}}$, as well as of the 
associated effective sound speed, lowers the critical amplitude required for 
collapse and therefore enhances the PBH formation probability on the mass scales 
corresponding to horizon entry during this transient 
epoch~\cite{Harada:2013epa}. In this respect, the mechanism at hand plays a role 
analogous to the QCD-induced softening discussed in standard PBH 
scenarios~\cite{Jedamzik:1996mr,Musco:2023dak}, with the important difference 
that here the effect arises purely from the gravitational sector and does not 
require any \emph{ad hoc} modification of the radiation fluid. Hence, the 
analysis presented below provides a concrete example of a transient 
gravitational mechanism for localized PBH enhancement within $f(T)$ cosmology.

The paper is organized as follows. In Sec. \ref{Sec.II} we briefly review the 
teleparallel framework and the cosmological field equations in $f(T)$ gravity, 
introducing the effective-fluid description that is convenient for the 
interpretation of the background dynamics. In Sec. \ref{Sec.III} we present the 
specific functional form of the $f(T)$ model under consideration, discuss its 
viability, and examine the resulting evolution of the effective torsion fraction 
and of the total EoS parameter. In Sec. \ref{Sec.IV} we investigate the 
implications for PBH formation in this time-dependent background, focusing on 
the collapse threshold and the resulting PBH mass function. Finally, in Sec. 
\ref{Sec.V} we summarize the main results and discuss possible extensions, 
including the comparison with current observational constraints on PBH 
abundances.

\section{The Teleparallel Framework}
\label{Sec.II}

In this section we briefly review the basic ingredients of teleparallel gravity 
and its cosmological application, setting the stage for the analysis that 
follows.

Teleparallel gravity is formulated in terms of the tetrad (vierbein) field 
$e^{A}{}_{\mu}$ and, in its covariant formulation, a spin connection that 
ensures local Lorentz covariance \cite{AldrovandiPereira2013,Krssak:2015oua}. 
The spacetime metric is constructed as
\begin{equation}
g_{\mu\nu}=\eta_{AB}\,e^{A}{}_{\mu}e^{B}{}_{\nu},\qquad 
\eta_{AB}=\mathrm{diag}(-1,+1,+1,+1),
\end{equation}
while $e\equiv \det(e^{A}{}_{\mu})=\sqrt{-g}$. 

In contrast to the Levi-Civita connection of general relativity, teleparallel 
gravity employs the Weitzenb\"ock connection, which is curvature-free but 
torsionful. In a convenient gauge, sufficient for the homogeneous and isotropic 
background considered here, it can be written as
\begin{equation}
\Gamma^{\rho}{}_{\mu\nu}=e_{A}{}^{\rho}\,\partial_{\nu} e^{A}{}_{\mu},
\end{equation}
leading to the torsion tensor
\begin{equation}
T^{\rho}{}_{\mu\nu}=\Gamma^{\rho}{}_{\nu\mu}-\Gamma^{\rho}{}_{\mu\nu}
=e_{A}{}^{\rho}\left(\partial_{\mu}e^{A}{}_{\nu}-\partial_{\nu}e^{A}{}_{\mu}
\right).
\end{equation}

Defining the contortion tensor $K^{\rho}{}_{\mu\nu}$ and the superpotential 
$S_{\rho}{}^{\mu\nu}$ as
\begin{align}
K^{\rho}{}_{\mu\nu} &\equiv 
-\frac{1}{2}\left(T^{\rho}{}_{\mu\nu}-T^{\rho}{}_{\nu\mu}-T_{\mu}{}^{\rho}{}_{
\nu}\right),\\
S_{\rho}{}^{\mu\nu} &\equiv \frac{1}{2}\left(K^{\mu\nu}{}_{\rho}
+\delta^{\mu}_{\rho}T^{\alpha\nu}{}_{\alpha}-\delta^{\nu}_{\rho}T^{\alpha\mu}{}_
{\alpha}\right),
\end{align}
one constructs the torsion scalar
\begin{equation}
T\equiv S_{\rho}{}^{\mu\nu}T^{\rho}{}_{\mu\nu}.
\end{equation}

For a spatially flat FLRW line element
\begin{equation}
\dd s^{2}=-\dd t^{2}+a(t)^{2}\dd\bm{x}^{2},\qquad \Hh\equiv \frac{\dot a}{a},
\end{equation}
and a diagonal tetrad $e^{A}{}_{\mu}=\mathrm{diag}(1,a,a,a)$, one obtains the 
simple relation
\begin{equation}
T=6\Hh^{2}>0.
\label{eq:T6H2}
\end{equation}
Note that the sign   of $T$ depends on the metric signature convention, hence 
  \eqref{eq:T6H2} is consistent with the metric signature adopted 
above and with standard cosmological treatments \cite{Cai:2015emx}.

We consider a minimally coupled $f(T)$ extension of the teleparallel equivalent 
of general relativity (TEGR), with action
\begin{equation}
S=\frac{1}{16\pi G}\int \dd^{4}x\, e\left[T+f(T)\right] + S_m[g_{\mu\nu},\psi],
\label{eq:action}
\end{equation}
where $S_m$ contains radiation and any additional matter fields coupled only to 
the metric. Denoting $f_T\equiv \dd f/\dd T$ and $f_{TT}\equiv \dd^{2}f/\dd 
T^{2}$, the modified Friedmann equations for a spatially flat FLRW background 
read \cite{Cai:2015emx,Krssak:2015oua}
\begin{align}
\Hh^{2} &= \frac{8\pi G}{3}\rh -\frac{f}{6}+\frac{T f_T}{3},\label{eq:Fried1}\\
\dot{\Hh} &= -\frac{4\pi G(\rh+p)}{1+f_T+2T f_{TT}}.\label{eq:Fried2}
\end{align}
Moreover, the minimal coupling ensures the standard conservation equation for 
the matter sector,
\begin{equation}
\dot{\rh}+3\Hh(\rh+p)=0.
\label{eq:consm}
\end{equation}

It proves particularly convenient to rewrite the system 
\eqref{eq:Fried1}-\eqref{eq:Fried2} in an effective two-fluid form, in which 
the modifications of gravity are encoded in an effective torsion fluid 
\cite{Cai:2015emx}. In this description one has
\begin{align}
3\Hh^{2} &= 8\pi G(\rh+\rhof),\label{eq:effF1}\\
-2\dot{\Hh} &= 8\pi G\left[(\rh+p)+(\rhof+\pf)\right],\label{eq:effF2}
\end{align}
where the effective torsion energy density and pressure are defined as
\begin{align}
\rhof &\equiv \frac{1}{16\pi G}\left(2T f_T-f\right),\label{eq:rhof}\\
\pf &\equiv \frac{1}{16\pi G}\left[f-2T f_T +4\dot{\Hh}(f_T+2T 
f_{TT})\right].\label{eq:pf}
\end{align}
Thus, the corresponding equation-of-state parameter is
\begin{equation}
\wf \equiv \frac{\pf}{\rhof}
= -1+\frac{4\dot{\Hh}(f_T+2T f_{TT})}{2T f_T-f}.
\label{eq:wf}
\end{equation}

Introducing the density parameters
\begin{equation}
\Omr\equiv \frac{8\pi G\rh_r}{3\Hh^{2}},\qquad
\Omf\equiv \frac{8\pi G\rhof}{3\Hh^{2}},
\end{equation}
and using the relation $T=6\Hh^{2}$, one obtains the algebraic expression
\begin{equation}
\Omf = \frac{2T f_T-f}{T}.
\label{eq:Omf_general}
\end{equation}
Hence, in the case of a radiation plus torsion-fluid system one has 
$\Omr+\Omf=1$. Furthermore, since the matter sector is conserved and the 
modified Friedmann equations form a closed system, the effective torsion 
component is separately conserved too, namely
\begin{equation}
\dot{\rhof}+3\Hh(\rhof+\pf)=0.
\label{eq:consf}
\end{equation}

\section{A model of transient $f(T)$ cosmology}
\label{Sec.III}

We now introduce the specific $f(T)$ model that will be used in the following 
analysis and examine its cosmological implications at the background level.

We begin by defining a positive reference torsion scale $T_\star \equiv 
6\Hh_\star^2>0$, and introducing the dimensionless variable
\begin{equation}
x \equiv \frac{T}{T_\star} = \frac{\Hh^2}{\Hh_\star^2}.
\end{equation}
In terms of this variable, we consider the $f(T)$ function
\begin{equation}
f(T) = \lambda T_\star x^{3} e^{-x}, \qquad x=\frac{T}{T_\star},
\label{eq:fTmodel}
\end{equation}
where $\lambda$ is a dimensionless coupling parameter.

The crucial property of the ansatz \eqref{eq:fTmodel} is that it induces a 
\emph{transient modification} of the cosmological dynamics. In particular, at 
late times ($T\to 0^+$) one has $f(T)\sim \lambda T_\star x^3 \to 0$, with 
$f_T=\mathcal{O}(x^2)\to 0$, while at early times ($T\to +\infty$) the 
exponential suppression ensures $f,f_T,f_{TT}\to 0$. Hence, the model 
asymptotically reduces to the teleparallel equivalent of general relativity 
both in the early- and late-time limits, while allowing for nontrivial 
deviations at intermediate epochs. 

It is worth noting that the functional form \eqref{eq:fTmodel} can be 
interpreted as a combination of power-law and exponential $f(T)$ models that 
have been shown to be consistent with cosmological observations, including 
supernovae type Ia, cosmic microwave background, baryon acoustic oscillations, 
and growth data~\cite{Nesseris:2013jea}.

From \eqref{eq:fTmodel} one readily obtains
\begin{equation}
f_T = \lambda x^2 (3-x)e^{-x}, \qquad
f_{TT} = \frac{\lambda}{T_\star}\,x(x^2-6x+6)e^{-x}.
\end{equation}
Substituting into \eqref{eq:Omf_general}, we obtain the explicit expression for 
the effective torsion fraction
\begin{equation}
\Omf(x) = \lambda x^2 e^{-x}(5-2x).
\label{eq:Omf_model}
\end{equation}

The condition $\wf=-1$ corresponds, from \eqref{eq:wf}, to $f_T+2T f_{TT}=0$. 
In the present model this yields
\begin{equation}
f_T+2T f_{TT} = \lambda x^2 e^{-x}\left(2x^2-13x+15\right),
\end{equation}
and therefore
\begin{equation}
\wf=-1 \quad \Longleftrightarrow \quad x\in\left\{\frac{3}{2},\,5\right\}.
\label{eq:wfminus1}
\end{equation}

We define a reference epoch $t_{\mathrm{ref}}$ by selecting the first crossing, 
i.e.
\begin{equation}
x(t_{\mathrm{ref}})=\frac{3}{2}.
\label{eq:xref}
\end{equation}
The coupling $\lambda$ is then fixed by requiring that the torsion sector 
contributes a specified, non-negligible fraction of the total energy density at 
$t_{\mathrm{ref}}$. Choosing
\begin{equation}
\lambda=\frac{1}{45}e^{3/2},
\label{eq:lambda}
\end{equation}
one obtains
\begin{equation}
\Omf\!\left(\frac{3}{2}\right)=0.1.
\label{eq:Omf01}
\end{equation}

At the reference epoch, the Universe consists of radiation and the effective 
torsion fluid, with $\Omr(t_{\mathrm{ref}})=0.9$ and $w_r=1/3$, while 
$\wf(t_{\mathrm{ref}})=-1$. The total equation-of-state parameter is thus
\begin{equation}
\wtot(t_{\mathrm{ref}})=\wf\Omf+w_r\Omr
=-0.1+\frac{1}{3}\times 0.9
=0.2,
\label{eq:wtotref}
\end{equation}
demonstrating explicitly a transient reduction relative to pure radiation. This 
deviation is entirely induced by the modified gravitational sector.

 \begin{figure}[ht]
  \centering
  \includegraphics[width=0.45\textwidth]{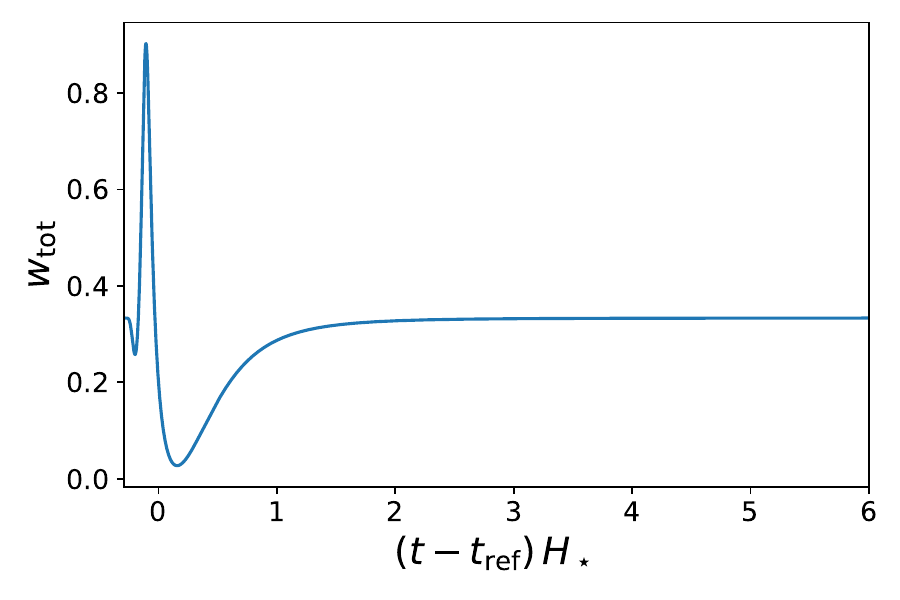}
\caption{\it 
Evolution of the total equation-of-state parameter $w_{\rm tot}(t)$ 
in the transient $f(T)$ model 
$f(T)=\lambda T_\star (T/T_\star)^3 e^{-T/T_\star}$ with $T=6H^2$ and 
$\lambda=\frac{0.2}{9}e^{3/2}$, as a function of the dimensionless shifted time 
$(t-t_{\mathrm{ref}})H_\star$. 
The reference epoch $t_{\mathrm{ref}}$ is defined by 
$x(t_{\mathrm{ref}})=T(t_{\mathrm{ref}})/T_\star=H(t_{\mathrm{ref}}
)^2/H_\star^2=3/2$, 
at which $\Omega_f(t_{\mathrm{ref}})=0.1$ and $w_{\rm 
tot}(t_{\mathrm{ref}})=0.2$. 
The figure illustrates the transient softening of the cosmic background 
relative to radiation domination ($w_{\rm tot}=1/3$), induced by the effective torsion 
sector.
}
\label{fig:w_of_t}
\end{figure}

 \begin{figure}[ht]
  \centering
  \includegraphics[width=0.45\textwidth]{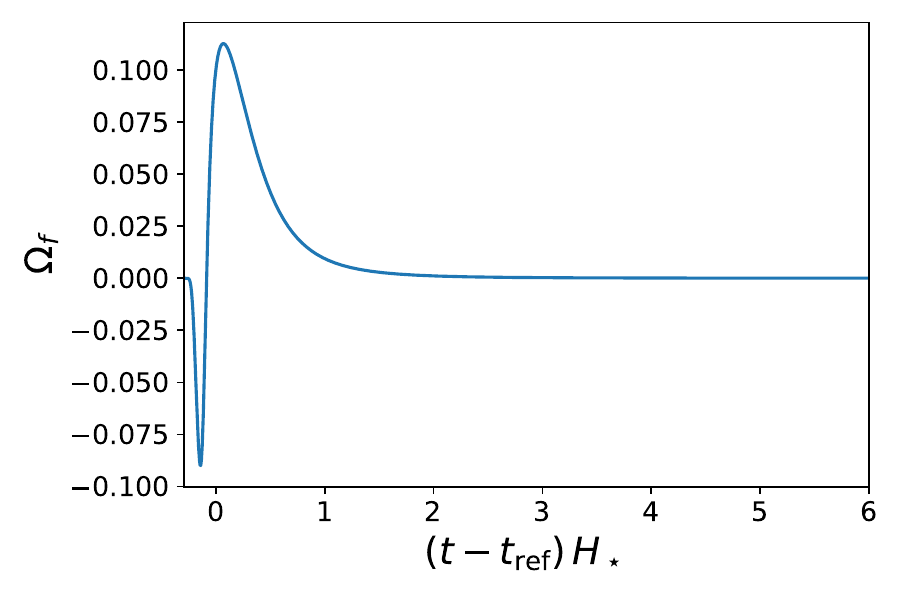}
\caption{\it 
Evolution of the effective torsion-fluid fraction $\Omega_f(t)$ for the 
transient $f(T)$ model, plotted as a function of the dimensionless shifted time 
$(t-t_{\mathrm{ref}})H_\star$, with $t_{\mathrm{ref}}$ defined by 
$x(t_{\mathrm{ref}})=3/2$. 
The coupling $\lambda=\frac{0.2}{9}e^{3/2}$ is fixed analytically so that 
$\Omega_f(t_{\mathrm{ref}})=0.1$. 
The figure clearly demonstrates the transient nature of the torsion sector, 
which becomes dynamically relevant around $t_{\mathrm{ref}}$, while being 
suppressed both at early times ($T/T_\star\to\infty$) and at late times 
($T/T_\star\to 0$).
}
\label{fig:Omega_f_of_t}
\end{figure}

The time evolution of $\wtot(t)$ and $\Omf(t)$ is obtained by integrating 
\eqref{eq:Fried2} together with the conservation equations \eqref{eq:consm} and 
\eqref{eq:consf}. Figs. \ref{fig:w_of_t} and~\ref{fig:Omega_f_of_t} 
illustrate the resulting background dynamics. In particular, the total EoS 
parameter departs from the radiation value $1/3$ within a finite interval 
around $t_{\mathrm{ref}}$, reaching $\wtot=0.2$, while returning to the 
standard radiation-dominated behavior outside this window. Physically, this 
reflects the fact that the torsion sector temporarily behaves as a 
vacuum-energy-like component, reducing the effective pressure without inducing 
accelerated expansion.

The corresponding evolution of $\Omega_f$ confirms that the torsion 
contribution is explicitly transient: it is suppressed both at early times 
($T/T_\star\to\infty$) and at late times ($T/T_\star\to 0$), while attaining a 
maximum contribution near $t_{\mathrm{ref}}$. The combined behavior of 
$\Omega_f$ and $w_{\mathrm{tot}}$ thus provides a clear realization of a 
time-localized modification of the cosmic background.

\begin{figure}[ht]
  \centering
  \includegraphics[width=0.45\textwidth]{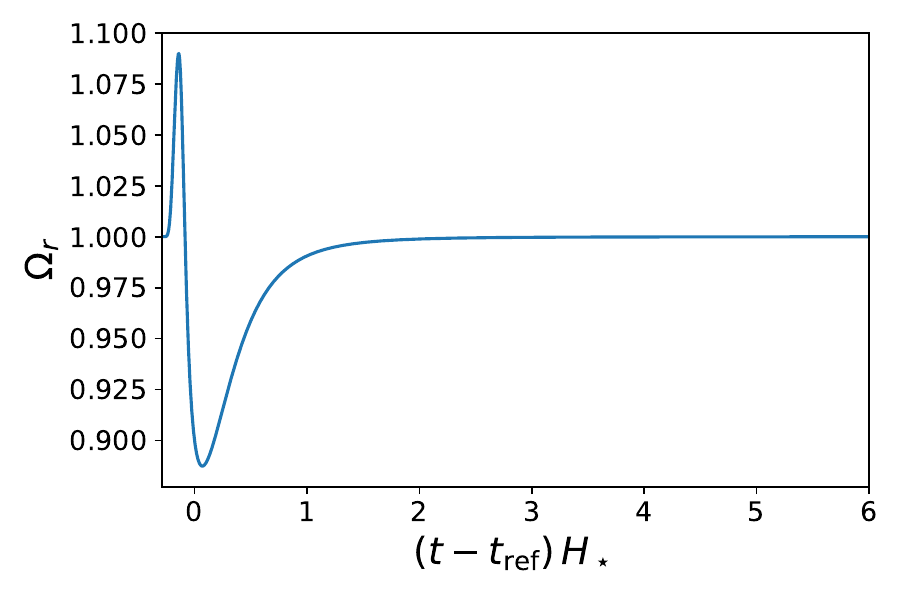}
\caption{\it 
Evolution of the radiation energy fraction $\Omega_r(t)$ as a function of the 
dimensionless shifted time $(t-t_{\mathrm{ref}})H_\star$ in the effective 
two-component (radiation + torsion) background. 
The evolution is obtained by integrating the background equations for $x\equiv 
H^2/H_\star^2$ with initial condition $x(t_{\mathrm{ref}})=3/2$. 
The figure illustrates the transient departure from pure radiation domination 
induced by the torsion sector, followed by a rapid restoration of $\Omega_r\to 
1$ as the modification becomes negligible.
}
\label{fig:omega_r_vs_tau}
\end{figure}

Fig. \ref{fig:omega_r_vs_tau} shows the evolution of the radiation fraction 
$\Omega_r$ around the reference epoch. Starting from $\Omega_r\simeq 0.9$ at 
$t_{\mathrm{ref}}$, it exhibits a mild dip and then monotonically approaches 
unity as the torsion component becomes negligible. This behavior captures a 
transient departure from radiation domination followed by a rapid restoration 
of the standard cosmological evolution.

It is important to note that a nontrivial extremum of $\Omf$ does not coincide 
with $\wf=-1$ in a minimally coupled two-component system. Indeed, using 
derivatives with respect to $N\equiv\ln a$, one finds
\begin{equation}
\Omf' = 3(w_r-\wf)\Omf(1-\Omf),
\label{eq:Omfprime}
\end{equation}
which implies that a turning point with $0<\Omf<1$ requires $\wf=w_r=1/3$. 
Consequently, the epoch where $\wf=-1$ is close to, but not identical with, the 
maximum of $\Omf$. In the present model the peak occurs at
\begin{equation}
x_{\mathrm{peak}}=\frac{11-\sqrt{41}}{4}\approx 1.1492,
\end{equation}
with $\Omf(x_{\mathrm{peak}})\approx 0.113$ for the choice \eqref{eq:lambda}.

To relate the reference epoch to a physical temperature, we employ radiation 
thermodynamics, namely we use
\begin{equation}
\rh_r(T)=\frac{\pi^2}{30}g_\star(T)T^4,
\end{equation}
together with \eqref{eq:effF1}, obtaining
\begin{equation}
\Hh_{\mathrm{ref}}^2=\frac{8\pi 
G}{3}\frac{1}{1-\Omf(t_{\mathrm{ref}})}\rh_r(T_{\mathrm{ref}}).
\label{eq:HvsT}
\end{equation}
The scale $T_\star$ is then fixed through $x(t_{\mathrm{ref}})=3/2$, yielding
\begin{equation}
T_\star=\frac{2}{3}T(t_{\mathrm{ref}}), \qquad 
\Hh_\star=\frac{\Hh_{\mathrm{ref}}}{\sqrt{3/2}}.
\end{equation}

In this work we focus on PBH formation at asteroid-mass scales, where current 
observational constraints are weak. Accordingly, we choose the horizon mass at 
the reference epoch to satisfy
\begin{equation}
M_H(t_{\rm ref}) \sim M_{\rm ast},
\end{equation}
ensuring that the relevant modes re-enter the horizon well before the QCD 
epoch, characterized by $T_{\rm QCD}\sim \mathcal{O}(0.1)\,\mathrm{GeV}$. 
Therefore, we impose
\begin{equation}
T_{\rm ref} > T_{\rm QCD},
\end{equation}
which sets the normalization of the temperature scale.
Additionally, using the background equations, one finds
\begin{equation}
x[1-\Omega_f(x)] = 1.35\left(\frac{T}{T_{\rm ref}}\right)^4,
\label{derivation}
\end{equation}
providing the mapping between the dimensionless variable $x$ and the physical 
temperature.

 \begin{figure}[ht]
  \centering
  \includegraphics[width=0.49\textwidth]{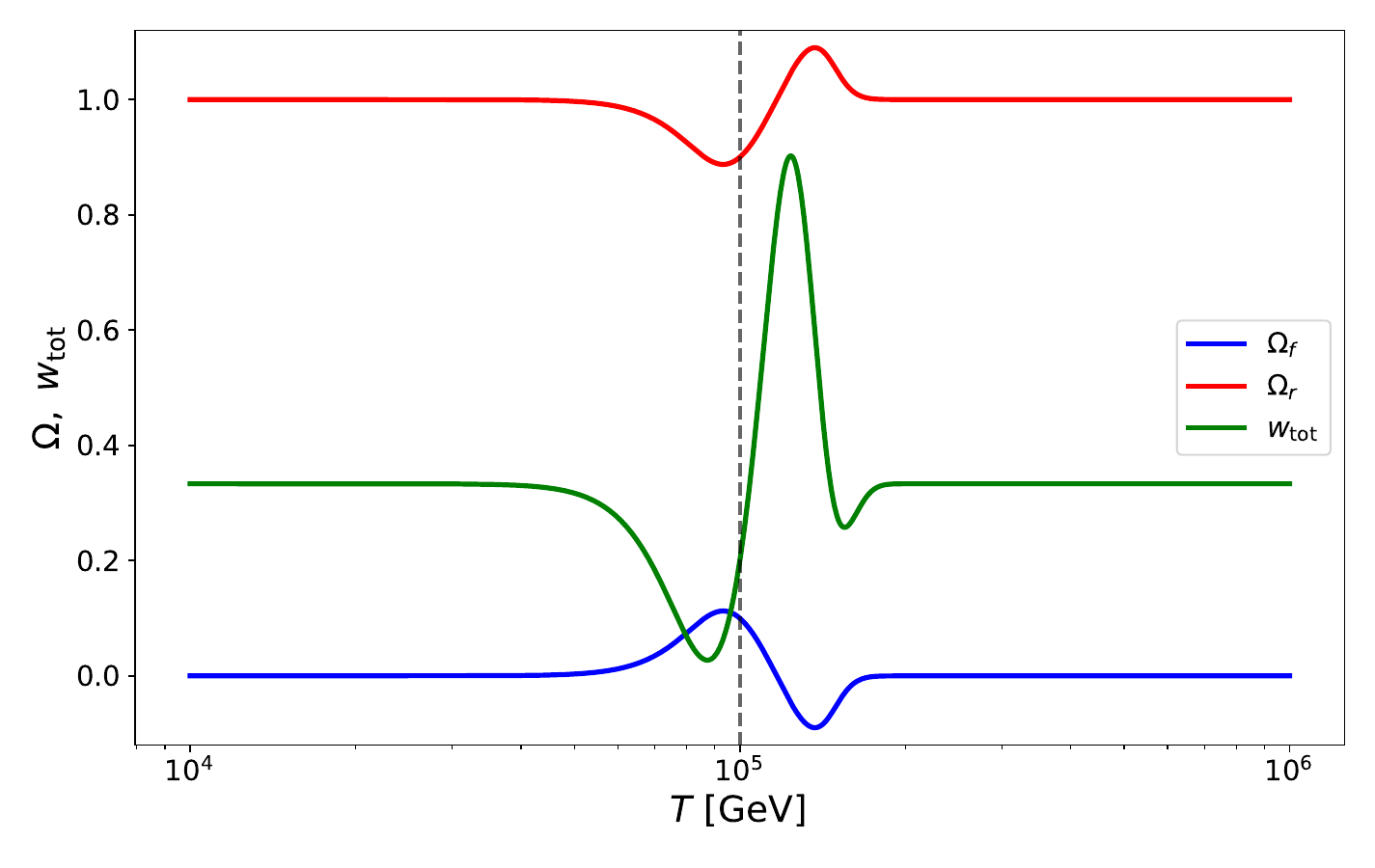}
\caption{\it 
Evolution of the effective torsion fraction $\Omega_f$ (blue), the radiation 
fraction $\Omega_r$ (red), and the total equation-of-state parameter $w_{\rm 
tot}$ (green), shown as functions of the primordial plasma temperature $T$. 
The transient $f(T)$ model is calibrated with $\lambda=e^{3/2}/45$, such that 
at the reference temperature $T_{\rm ref}=10^{5}\,\mathrm{GeV}$ (vertical 
dashed line) one has $x(T_{\rm ref})=3/2$, yielding $\Omega_f(T_{\rm ref})=0.1$ 
and $w_{\rm tot}(T_{\rm ref})=0.2$. 
The figure illustrates the time-localized impact of the torsion sector, which 
induces a transient reduction of the total EoS while the radiation component 
remains dominant, before the system returns to standard radiation-dominated 
behavior at both higher and lower temperatures.
}
\label{fig:Omega_w_vs_T}
\end{figure}

Finally, Fig.~\ref{fig:Omega_w_vs_T} summarizes the combined evolution of 
$\Omega_r$, $\Omega_f$, and $w_{\mathrm{tot}}$. We note that 
$\Omega_f(x)\propto (5-2x)$ changes sign for $x>5/2$, leading to a brief 
negative excursion. This does not signal any pathology, since $\Omega_f$ 
represents an effective geometric contribution, and in modified gravity, 
individual effective components are not required to lie between $0$ and $1$, 
provided that the total density parameter satisfies $\Omega_{\mathrm{tot}}=1$.

\section{Primordial black hole formation}
\label{Sec.IV}

In this section we investigate the implications of the transient $f(T)$ 
cosmological dynamics for primordial black hole (PBH) formation. In particular, 
we focus on how the time-localized modification of the background equation of 
state affects the collapse threshold and, consequently, the resulting PBH 
abundance.

\subsection{Threshold for PBH formation}

Primordial black holes form when an overdense region collapses against pressure 
gradients and cosmological expansion after horizon re-entry. The collapse 
process is typically characterized in terms of the energy density contrast 
smoothed on a scale $R$, denoted by $\delta_R$, or equivalently through the 
compaction function. Collapse occurs when the perturbation amplitude exceeds a 
critical threshold $\delta_c$, which depends on the background equation of 
state (EoS)~\cite{Carr:1975qj} as well as on the shape of the perturbation 
profile~\cite{Musco:2018rwt}. 

For a perfect fluid with $p=w_{\rm tot}\rh$ and $w_{\rm tot}\geq 0$, both analytic arguments and 
numerical simulations indicate that the collapse threshold $\delta_c$ increases 
with $w_{\rm tot}$. In particular, for radiation domination one finds 
$\delta_c=\mathcal{O}(0.4)$~\cite{Harada:2013epa,Musco:2012au}. This reflects 
the fact that larger pressure gradients more efficiently counteract 
gravitational collapse.

In the present $f(T)$ scenario, the background during the transient epoch is 
not described by a single barotropic fluid, but rather by two separately 
conserved components: radiation and an effective torsion fluid with 
time-dependent EoS parameter $\wf(t)$. The total pressure and energy density are 
therefore
\begin{equation}
p_{\mathrm{tot}}=p_r+\pf=w_r\rh_r+\wf\rhof,
\qquad
\rh_{\mathrm{tot}}=\rh_r+\rhof,
\end{equation}
leading to the total EoS parameter
\begin{equation}
\wtot(t)=\frac{p_{\mathrm{tot}}}{\rh_{\mathrm{tot}}}
=w_r\Omr(t)+\wf(t)\Omf(t).
\label{eq:wtot}
\end{equation}

For the parameter choice \eqref{eq:lambda}, at the reference epoch 
\eqref{eq:xref} one obtains $\wtot(t_{\mathrm{ref}})=0.2$, as shown in 
\eqref{eq:wtotref}. This corresponds to a transient softening of the background 
relative to the standard radiation value $1/3$. Physically, the EoS determines 
the efficiency of pressure gradients and the associated propagation of sound 
waves across the collapsing region. A reduction in $\wtot$, and more generally 
in the adiabatic sound speed $c_a^2\equiv \dot p_{\mathrm{tot}}/\dot 
\rh_{\mathrm{tot}}$, decreases the efficiency of pressure support and thus 
facilitates gravitational collapse.

To obtain a quantitative estimate of this effect, one may employ the analytic 
expression of Harada, Kohri and Yoo~\cite{Harada:2013epa}, which provides an 
estimate of the collapse threshold in comoving slicing as a function of the total EoS 
parameter $w_{\rm tot}$:
\begin{equation}
\delta_c(w_{\rm tot})=\frac{3(1+w_{\rm tot})}{5+3w_{\rm tot}}\,
\sin^{2}\!\left(\frac{\pi\sqrt{w_{\rm tot}}}{1+3w_{\rm tot}}\right).
\label{eq:deltaC}
\end{equation}

\begin{figure}[ht]
  \centering
\includegraphics[width=0.47\textwidth]{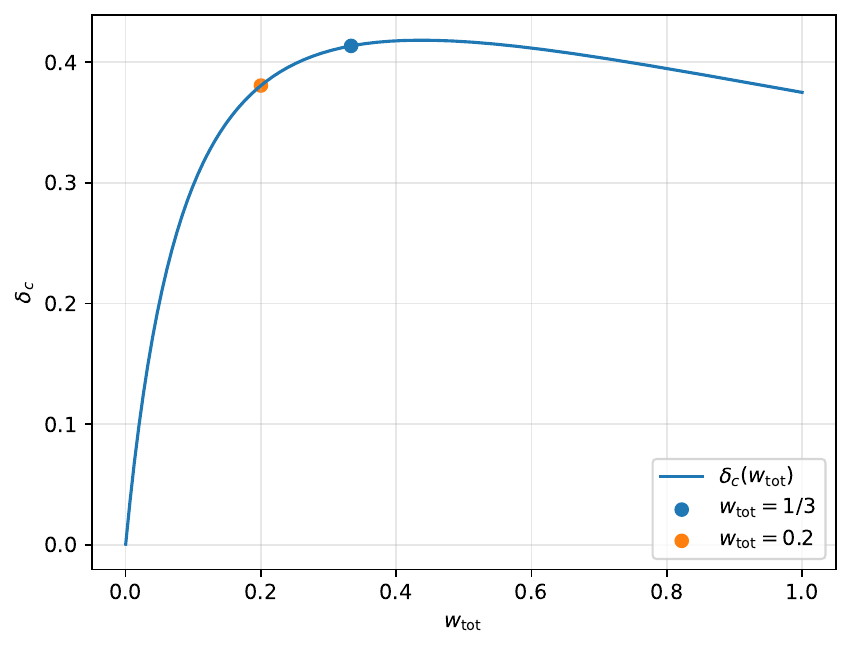}
\caption{\it 
The PBH formation threshold $\delta_c$ as a function of the total 
equation-of-state 
parameter $w_{\rm tot}$, based on the analytic estimate of Harada-Kohri-Yoo. 
The curve illustrates the strong dependence of the collapse threshold on the 
background EoS, with larger pressure (higher $w_{\rm tot}$) leading to 
increased 
resistance to gravitational collapse. 
The markers indicate representative values for radiation domination 
($w_{\rm tot}=\tfrac{1}{3}$) and for a softened background ($w_{\rm tot}=0.2$), 
highlighting how 
even a modest reduction of $w_{\rm tot}$ leads to a decrease of $\delta_c$ and 
thus 
enhances PBH formation.
}
\label{fig:deltac_vs_w}
\end{figure}

Fig. \ref{fig:deltac_vs_w} displays the corresponding dependence of 
$\delta_c$ on $w_{\rm tot}$. As $w_{\rm tot}$ increases, pressure gradients become more efficient 
and the threshold rises, reaching a broad maximum before decreasing for larger 
values of $w_{\rm tot}$~\footnote{According to recent numerical 
simulations~\cite{Escriva:2020tak}, the PBH formation threshold continues to 
increase for $w_{\rm tot}>1/3$. Equation \eqref{eq:deltaC} should therefore be viewed as a 
lower-bound estimate, valid primarily for $w_{\rm tot}<1/3$.}. The reference values at 
$w_{\rm tot}=1/3$ and $w_{\rm tot}=0.2$ illustrate how even a modest reduction of the effective EoS 
leads to a decrease of the collapse threshold.

Although the two-component system arising in $f(T)$ gravity does not correspond 
exactly to a single fluid with constant $w_{\rm tot}$, the relation \eqref{eq:deltaC} 
provides a reliable first estimate of the sensitivity of $\delta_c$ to the 
effective background EoS~\cite{Papanikolaou:2022cvo}. In particular, one finds 
$\delta_c\simeq 0.41$ for $w_{\rm tot}=1/3$ and $\delta_c\simeq 0.38$ for $w_{\rm tot}=0.2$. Even 
such a small reduction is significant, since the PBH abundance depends 
exponentially on $\delta_c$.

It is important to emphasize, however, that the collapse threshold depends in 
general on additional physical effects, including 
non-sphericities~\cite{Escriva:2024aeo,Escriva:2024lmm}, the shape of the 
perturbation profile~\cite{Musco:2018rwt,Escriva:2019phb}, and possible 
non-Gaussianities~\cite{Kehagias:2019eil,Atal:2019erb}. In the present analysis 
we do not incorporate these effects, as our primary aim is to isolate and 
quantify the impact of the modified-gravity-induced reduction of the effective 
EoS.

We can now summarize the physical mechanism underlying PBH formation in the 
present framework. The $f(T)$ model \eqref{eq:fTmodel} generates a transient, 
parameter-controlled departure from radiation domination, during which the 
effective torsion component becomes non-negligible and $\wf$ approaches 
negative values near the first crossing $\wf=-1$ at $x=3/2$. This leads to a 
temporary reduction of $\wtot$ from $1/3$ to $0.2$, as given in 
\eqref{eq:wtotref}, and consequently to a suppression of pressure gradients at 
horizon re-entry. 

As a result, for primordial perturbations close to the collapse threshold, the 
reduction in $\delta_c$ enhances the PBH formation probability over a narrow 
mass range centered around the horizon mass at the reference epoch. In this 
sense, the torsion sector plays a role analogous to the QCD-induced softening 
in standard PBH scenarios, while remaining negligible outside the transient 
window and thus preserving the standard cosmological evolution at early and late 
times.

\subsection{The PBH mass function}

We now proceed to the calculation of the PBH abundance and mass function 
induced by the transient modification of the background dynamics. As we have 
seen above, the key effect of the $f(T)$ sector is to reduce the effective 
collapse threshold over a finite range of horizon masses. This effect can be 
translated into the PBH mass function once the statistics of the primordial 
fluctuations and the relation between horizon mass and PBH mass are specified.

Working within the Press-Schechter formalism~\cite{1974ApJ...187..425P}, and 
assuming nearly Gaussian statistics for the energy-density fluctuations at 
horizon entry, the probability density function (PDF) of the density contrast 
$\delta$ is taken to be
\begin{equation}
P(\delta)=\frac{1}{\sqrt{2\pi\sigma^2(M_H)}}\exp\!\left(-\frac{\delta^2}{
2\sigma^2(M_H)}\right),
\label{eq:gauss_pdf}
\end{equation}
where $\sigma^2(M_H)$ denotes the variance of the smoothed density field on the 
relevant horizon scale. The mass fraction of the Universe collapsing into PBHs 
at formation is then
\begin{align}
\beta_\mathrm{PS}(M) &\simeq 
2\int_{\delta_c}^{\infty}\frac{1}{\sqrt{2\pi}\sigma(M)}
\exp\!\left(-\frac{\delta^{2}}{2\sigma(M)^{2}}\right)\dd\delta
\nonumber \\
&=\mathrm{erfc}\!\left(\frac{\delta_c}{\sqrt{2}\,\sigma(M)}\right),
\label{eq:beta}
\end{align}
where the factor of $2$ accounts for the usual cloud-in-cloud correction, 
namely for under-threshold regions embedded inside larger collapsing 
regions~\cite{Jedamzik:1994nr}. The variance $\sigma^2(M)$ is related to the 
energy-density power spectrum $\mathcal{P}_\delta(k)$ smoothed over scales 
around horizon entry~\cite{Ando:2018qdb,Young:2019osy}. In the present 
analysis, and in order to isolate clearly the effect of the modified-gravity 
background, we assume for simplicity an approximately scale-invariant 
$\mathcal{P}_\delta(k)$ over the limited range of scales relevant to the 
transient epoch, leading to an approximately constant value of $\sigma^2$, 
which will be treated as a free parameter. Since $\beta$ depends on the ratio 
$\delta_c/\sigma$ through the complementary error function, even a modest 
reduction of $\delta_c$ can lead to an exponentially enhanced PBH abundance.

The characteristic PBH mass is set by the horizon mass at the time of collapse. 
In a flat FLRW universe one has
\begin{equation}
M_H \equiv \frac{4\pi}{3}\rh_{\mathrm{tot}}\Hh^{-3}
=\frac{4\pi}{3}\left(\frac{3\Hh^{2}}{8\pi G}\right)\Hh^{-3}
=\frac{1}{2G\Hh}
=\frac{4\pi \mpl^{2}}{\Hh},
\label{eq:horizonmass}
\end{equation}
while the PBH mass is typically of the order
\begin{equation}
M_{\mathrm{PBH}}\simeq \gamma M_H,
\end{equation}
with $\gamma=\mathcal{O}(0.1)$, depending on the details of the collapse and in 
particular on the equation of state at the time of 
formation~\cite{Maison:1995cc,Musco:2012au}.

The relation between temperature and Hubble rate in the present scenario 
differs from the standard radiation-dominated one due to the torsion 
contribution. In particular, from \eqref{eq:HvsT} we obtain
\begin{equation}
H^2(T)=\frac{8\pi G}{3}\rho_{\rm tot}(T)
=\frac{8\pi G}{3}\frac{\rho_r(T)}{1-\Omega_f(T)},
\label{eq:H_fT}
\end{equation}
and therefore
\begin{equation}
M_H(T)=\frac{1}{2G H(T)}=\sqrt{1-\Omega_f(T)}\,M_H^{\rm GR}(T),
\label{eq:MH_fT}
\end{equation}
where the standard GR expression is
\begin{equation}
M_H^{\rm GR}(T)\simeq 1.5\times 10^{5}M_\odot
\left(\frac{g_\star(T)}{10.75}\right)^{-1/2}
\left(\frac{T}{\mathrm{MeV}}\right)^{-2}.
\label{eq:MH_GR_T}
\end{equation}
Hence, at fixed temperature, a positive torsion contribution $\Omega_f>0$ 
increases the Hubble rate and correspondingly decreases the horizon mass by a 
factor $\sqrt{1-\Omega_f}$.

A more accurate description of PBH formation must also take into account the 
critical-collapse scaling 
law~\cite{Niemeyer:1997mt,Niemeyer:1999ak,Musco:2008hv,Musco:2012au}, according 
to which PBHs do not form with exactly the horizon mass. Instead, their mass 
obeys
\begin{equation}
\label{eq:PBH_mass_scaling_law}
M_\mathrm{PBH} = M_\mathrm{H}\mathcal{K}(\delta-\delta_\mathrm{c})^\gamma,
\end{equation}
where $M_\mathrm{H}$ is given by \eqref{eq:MH_fT}, $\gamma$ is the critical 
exponent, and $\mathcal{K}$ is a coefficient that depends on the EoS and on the 
shape of the collapsing profile. In the present work we take $\gamma=0.36$, 
which is the standard value for PBH formation during radiation domination, and 
adopt the representative choice $\mathcal{K}\simeq 4$~\cite{Musco:2008hv}. 
Under this refinement, the PBH mass fraction becomes
\begin{align}\label{eq:beta_M_corrected}
\beta(M) &\simeq 
2\int_{\delta_c}^{\infty}\frac{M}{M_\mathrm{H}}P(\delta)\dd\delta 
\nonumber \\
&= 
2\int_{\delta_c}^{\infty}\mathcal{K}(\delta-\delta_\mathrm{c})^\gamma\frac{1}{
\sqrt{2\pi}\sigma(M)}
\exp\!\left(-\frac{\delta^{2}}{2\sigma(M)^{2}}\right)\dd\delta.
\end{align}

Since PBHs behave as pressureless matter after formation, their fractional 
contribution grows relative to the radiation background until matter-radiation 
equality. Therefore, their mass fraction at equality is
\begin{equation}
\beta_{\rm eq}(M)
=\frac{a_{\rm eq}}{a(M)}\,\beta(M)
=\left(\frac{M_{\rm eq}}{M}\right)^{1/2}\beta(M),
\label{eq:beta_eq}
\end{equation}
where $M_{\rm eq}\simeq 2.8\times 10^{17}M_\odot$ is the horizon mass at 
equality. After equality, the PBH fraction remains approximately constant, and 
thus the present-day fraction of cold dark matter in PBHs per logarithmic mass 
interval is
\begin{equation}
f_\mathrm{PBH}(M)\equiv \frac{1}{\Omega_{\rm CDM}}\frac{d\Omega_{\rm PBH}}{d\ln 
M}
\simeq \frac{1}{\Omega_{\rm CDM}}\left(\frac{M}{M_{\rm 
eq}}\right)^{-1/2}\beta(M).
\label{eq:f_mono}
\end{equation}

Finally, by changing the integration variable in \eqref{eq:beta_M_corrected} 
from $\delta$ to $M_H$, one obtains the extended PBH mass function in the 
form~\cite{Byrnes:2018clq}
\begin{align}
f_\mathrm{PBH}(M)=& \frac{1}{\Omega_{\rm CDM}}\int_{-\infty}^{M_{\rm eq}} d\ln 
M_H\;
\frac{2}{\sqrt{2\pi\sigma^2(M_H)}} \nonumber \\
\times &\exp\!\left[
-\frac{\bigl(\mu^{1/\gamma}+\delta_c(M_H)\bigr)^2}{2\sigma^2(M_H)}
\right]
\frac{M}{\gamma M_H}\,\mu^{1/\gamma}\,\left(\frac{M_{\rm 
eq}}{M_H}\right)^{1/2}, \nonumber \\
\mu\equiv &\frac{M}{\mathcal{K}M_H}.
\label{eq:f_extended}
\end{align}
\begin{figure}[ht]
  \centering
  \includegraphics[width=0.49\textwidth]{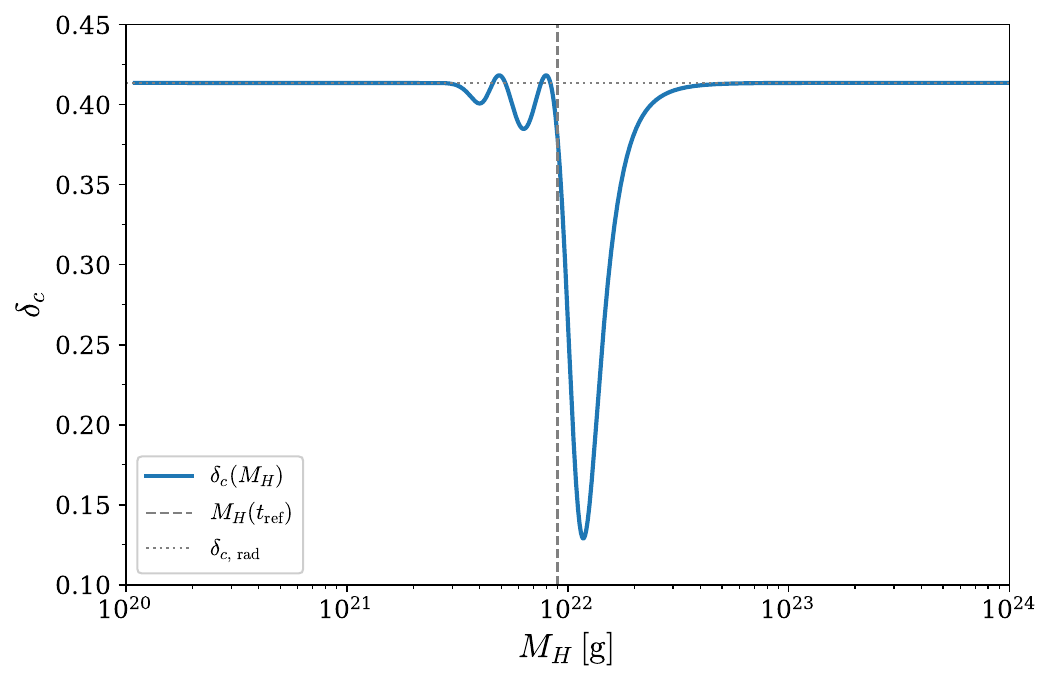}
\caption{\it 
The PBH formation threshold $\delta_c(M_H)$ as a function of the horizon mass 
$M_H$ in the transient $f(T)$ scenario. 
The dashed vertical line marks the reference mass scale $M_H(t_{\rm ref})$, 
corresponding to $T_{\rm ref}=10^5\,\mathrm{GeV}$ and $\Omega_f(t_{\rm 
ref})=0.1$, while the dotted horizontal line indicates the standard radiation 
value $\delta_{c,\rm rad}=0.41354$. 
The figure demonstrates the localized reduction of the collapse threshold 
around the transient epoch, reflecting the temporary softening of the background 
and leading to an enhanced PBH formation probability over a narrow mass range.
}
\label{fig:dc_vs_MH}
\end{figure}

In order to visualize how the transient departure from pure radiation 
domination affects PBH formation across scales, it is useful to translate the 
time dependence of the background into a mass-dependent collapse threshold. For 
each epoch, or equivalently for each value of the dimensionless variable 
$x\equiv H^2/H_\star^2$, we compute the total EoS parameter $w_{\rm tot}(x)$ for 
the combined radiation and effective torsion sectors and then evaluate the 
corresponding threshold $\delta_c$ through \eqref{eq:deltaC}. Mapping the 
result to the horizon mass using \eqref{eq:MH_fT}, one obtains the function 
$\delta_c(M_H)$.

Fig. \ref{fig:dc_vs_MH} shows that $\delta_c(M_H)$ approaches the standard 
radiation value $\delta_{c,\rm rad}$ away from the transient epoch, while it is 
reduced in the vicinity of the reference horizon-mass scale determined by 
$T_{\rm ref}$. In particular, at $t_{\rm ref}$ the calibration $\Omega_f(t_{\rm 
ref})=0.1$ implies $w_{\rm tot}(t_{\rm ref})=0.2$ and therefore 
$\delta_c(t_{\rm ref})\simeq 0.381$. This already indicates that a modest 
softening of the background can lead to a sizable enhancement of the PBH 
abundance. We stress, however, that the minimum of $\delta_c(M_H)$ does not 
coincide exactly with the reference epoch. Although $\delta_c(t_{\rm ref})\simeq 
0.381$, the actual minimum is reached at a nearby horizon mass, where 
Fig.~\ref{fig:dc_vs_MH} gives $\delta_{c,\min}(M_H)\approx 0.12$.

\begin{figure}[ht]
  \centering
  \includegraphics[width=0.5\textwidth]{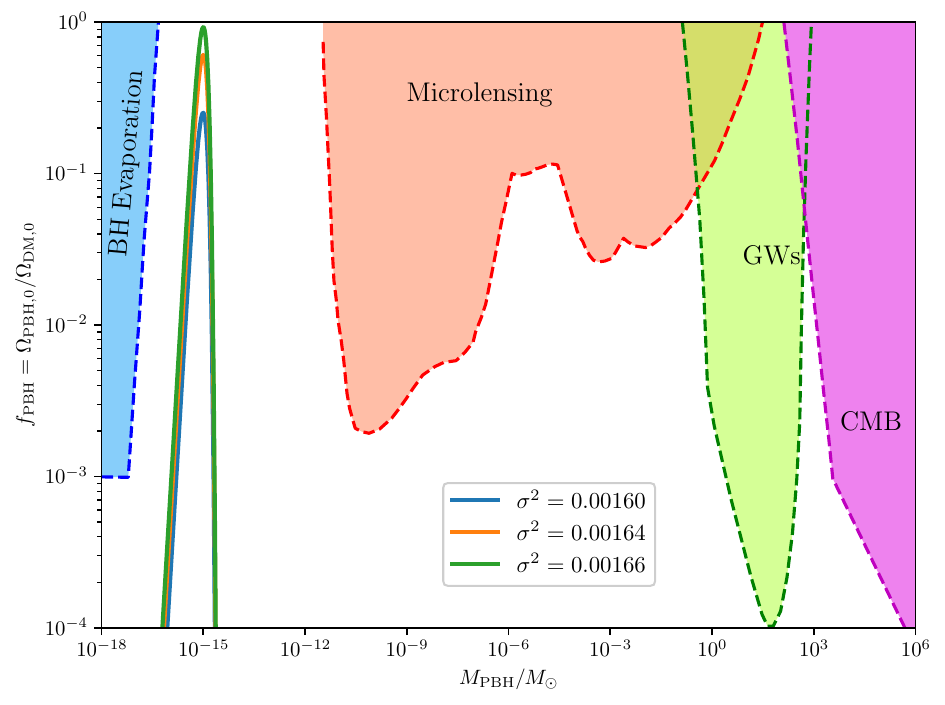}\caption{\it 
Constraints on the present-day PBH dark-matter fraction $f_{\rm PBH}(M)\equiv 
d\Omega_{\rm PBH}/d\ln M\,/\,\Omega_{\rm DM,0}$ as a function of the PBH mass 
$M_{\rm PBH}$. 
The shaded regions indicate excluded parameter space from BH evaporation 
(blue), microlensing (orange), stochastic gravitational-wave backgrounds 
(green), and CMB spectral distortions (magenta), with the corresponding 
exclusion boundaries shown as dashed curves. 
Superimposed are illustrative PBH mass functions $f(M)$ computed from 
Eq.~\eqref{eq:f_extended}. 
The left set of curves corresponds to distributions peaked at $M_{\rm peak}\sim 
10^{-15}M_\odot$ for different choices of the variance $\sigma^2$, while the 
right set corresponds to distributions peaked at $M_{\rm peak}\sim 1\,M_\odot$ 
with $\max f(M)=10^{-3}$. 
These examples demonstrate how the transient $f(T)$ mechanism can generate 
sharply peaked PBH mass functions that can account for a significant fraction 
of dark matter while remaining compatible with current observational 
constraints.
}
  \label{fig:pbh_abundance}
\end{figure}

Fig. \ref{fig:pbh_abundance} presents the resulting PBH dark-matter fraction 
$f_\mathrm{PBH}(M)$ computed from \eqref{eq:f_extended}, superimposed on 
current 
observational constraints as a function of the PBH mass $M_{\rm PBH}$ in 
solar-mass units. The shaded regions correspond to exclusions from BH 
evaporation at very low 
masses~\cite{Laha:2019ssq,Laha:2020ivk,Dasgupta:2019cae}, microlensing at 
intermediate masses~\cite{Niikura:2017zjd}, stochastic gravitational-wave 
backgrounds~\cite{Sasaki:2016jop,Andres-Carcasona:202405732}, and CMB spectral 
distortions~\cite{Ali-Haimoud:2016mbv}.

On top of these constraints we display illustrative families of curves obtained 
from \eqref{eq:f_extended} by choosing the parameters so that the PBH abundance 
is sharply peaked around the asteroid-mass scale, $M_{\rm peak}\sim 
10^{-15}M_\odot$, with $\max f(M)\lesssim 1$. These examples show that 
relatively modest changes in the collapse statistics can shift both the peak 
mass and the overall normalization of the PBH distribution, while remaining 
consistent with the current observational bounds.

\section{Conclusions}
\label{Sec.V}

In this work we have shown that a specific minimally coupled $f(T)$ 
teleparallel model can naturally give rise to a \emph{transient} effective 
torsion component, which is negligible at very early and very late times, but 
becomes dynamically relevant over a finite intermediate epoch. For the explicit 
analytic form in Eq.~\eqref{eq:fTmodel}, the effective torsion fraction 
$\Omega_f$ is given by Eq.~\eqref{eq:Omf_model}, and the model admits epochs 
where $w_f=-1$ at $x\in\{3/2,\,5\}$. By fixing the model parameters so that 
$\Omega_f$ attains a phenomenologically relevant value at the first $w_f=-1$ 
crossing, the coupling $\lambda$ is determined as in \eqref{eq:lambda}, leading 
to a temporary reduction of the total equation-of-state parameter to $w_{\rm 
tot}\simeq 0.2$ at the reference epoch. At the same time, the radiation 
component remains dominant, with $\Omega_r(t)$ deviating only mildly from unity 
and rapidly returning to the standard radiation-dominated behavior outside the 
transient window, ensuring consistency with the conventional cosmological 
evolution.

The implications for primordial black hole formation follow directly from the 
dependence of the collapse threshold on the background equation of state. Using 
the Harada-Kohri-Yoo estimate \eqref{eq:deltaC}, the threshold $\delta_c(w_{\rm tot})$ 
decreases as the effective EoS is reduced. Even a modest $\mathcal{O}(10\%)$ 
change in $\delta_c$ is sufficient to produce an exponentially enhanced PBH 
abundance for nearly Gaussian fluctuations, due to the sensitivity of the 
formation fraction to the ratio $\delta_c/\sigma$.

Incorporating the critical-collapse scaling of the PBH mass spectrum, and 
adopting a nearly scale-invariant density-contrast variance over the 
horizon-mass interval associated with the transient epoch, we find that the PBH 
abundance can be normalized to account for a significant fraction, or even the 
entirety, of the cold dark matter for values of the variance of order 
$\sigma^2\sim \mathcal{O}(10^{-3})$. The resulting PBH population is sharply 
peaked around the horizon mass corresponding to the transient epoch (modulo the 
expected $\mathcal{O}(1)$ shift due to critical collapse), with most of the PBH 
energy density concentrated within approximately one order of magnitude in 
mass. 

The above results demonstrate that the teleparallel torsion sector can provide 
a mechanism for PBH production that is closely analogous to the QCD-induced 
softening of the equation of state in standard scenarios, but with the 
important distinction that here the effect arises purely from the gravitational 
sector and is confined to a controlled transient epoch. In particular, the 
enhancement of PBH formation does not rely on any \emph{ad hoc} modification of 
the radiation fluid. Furthermore, varying the normalization of $\Omega_f$ at the 
first $w_f=-1$ crossing generically leads to a corresponding change in the 
collapse threshold, indicating that the enhancement mechanism is robust and does 
not require fine tuning of the model parameters.

Several extensions would further refine the phenomenology. A fully consistent 
treatment would require evolving perturbations through the transient 
two-component background and determining the collapse threshold using dedicated 
numerical simulations in this setting. In addition, it would be important to 
assess the impact of profile dependence, non-sphericity, and non-Gaussianity, 
as well as to confront the resulting extended PBH mass function with current 
observational constraints and merger-rate data. More broadly, it would be 
interesting to investigate the extent to which similar localized PBH features 
arise in other classes of transient $f(T)$ models, and whether they can be 
correlated with independent cosmological signatures of the same torsion-driven 
epoch. These investigations lie beyond the scope of the present work and are 
left for future projects.

\section*{Acknowledgments}
The authors would like to acknowledge the contribution of the LISA Cosmology 
Working Group (CosWG). They acknowledge as well support from the COST  Actions 
CA21136 -  Addressing observational tensions in cosmology with  systematics and 
fundamental physics (CosmoVerse)  - CA23130, Bridging high and low energies in 
search of quantum gravity (BridgeQG)  and CA21106 - COSMIC WISPers in the Dark 
Universe: Theory, astrophysics and  experiments (CosmicWISPers). Theodoros Papanikolaou receives financial support by the funding program 
“MEDICUS”  of the University of Patras.
\bibliography{references}   

\end{document}